 \documentclass[preprint2]{aastex}

\newcommand{\2}{{~\sc ii}}

\newcommand{\mic}{{\,$\mu$m}}

\shorttitle{Advanced optimal extraction for Spitzer/IRS}
\shortauthors{Lebouteiller et al.}

\begin{document}


\title{Advanced optimal extraction for the Spitzer/IRS}


\author{V.\ Lebouteiller, J.\ Bernard-Salas, G.C.\ Sloan, and D.J.\ Barry}
\affil{Center for Radiophysics and Space Research, Cornell University, Space Sciences Building, Ithaca, NY 14853-6801, USA}
\email{vianney@isc.astro.cornell.edu}







\begin{abstract}
We present new advances in the spectral extraction of point-like sources adapted to the \textit{Infrared Spectrograph} onboard the \textit{Spitzer Space Telescope}. For the first time, we created a super-sampled point spread function of the low-resolution modules. We describe how to use the point spread function to perform optimal extraction of a single source and of multiple sources within the slit. We also examine the case of the optimal extraction of one or several sources with a complex background. The new algorithms are gathered in a plugin called \texttt{AdOpt} which is part of the \texttt{SMART} data analysis software.
\end{abstract}





\section{Introduction}

\begin{figure}[b!]
\epsscale{1.0}
\plotone{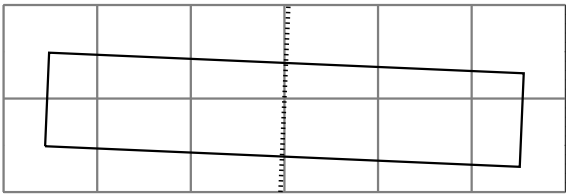}
\caption{The standard extraction aperture for a single wavelength, superimposed on the grid defined by the pixels on the detector array. The quadrilateral represents the extraction window, which is a ``tapered-column" whose width increases proportionally with wavelength to account for the varying point spread function. The tilt of the quadrilateral reflects the fact that the row axis of the detector array is not parallel to a line of constant wavelength.\label{fig:pseudo}}
\end{figure}

The ideal spectral extraction algorithm for point-like sources yields the maximum signal-to-noise ratio (S/N) while at the same time preserving the spectrophotometric fidelity. The standard extraction method is based on co-adding the flux in the cross-dispersion direction within a window large enough to contain (most of) the source flux. This method is known as ``tapered-column" in the  ``Spectroscopy Modeling Analysis and Reduction Tool" (\texttt{SMART}\footnote{\texttt{SMART} is available at \textit{http://isc.astro.cornell.edu/smart/}.}, Higdon et al.\ 2004) and is equivalent to the ``regular extraction" of the \textit{Spitzer} IRS Custom Extractor (\texttt{SPICE}), provided by the \textit{Spitzer Science Center} (\textit{SSC}). Although this generally produces satisfactory results, the inclusion of noisy pixels which do not contain a significant fraction of the flux inevitably tends to degrade the quality of the extracted spectrum. This is because every pixel in the extraction window is given the same weight. 
Moreover, the extraction window is part of a pseudo-rectangle which is defined as a zone in the detector array where the wavelength is uniform (Figure\,\ref{fig:pseudo}). When a quadrilateral boundary crosses a pixel, the signal is assumed to be evenly distributed within that pixel, which can lead to artificial flux variations in the extracted spectrum. Because of the interplay of the angled spectral trace and the widening extraction aperture, this error oscillates with wavelength with an amplitude which decreases toward longer wavelengths.

Knowledge of the point spread function (PSF) of the instrument can solve these problems, as it enables the so-called optimal extraction technique which weights the extracted data by the S/N of each pixel (Horne 1986). Optimal extraction therefore significantly reduces the statistical noise in the final spectrum compared to more typical extraction algorithms, which weight all the data within the window equally. So far, efforts on optimal extraction for the \textit{IRS} have made use of template PSFs or analytical PSFs to fit the cross-dispersion profile of the data:
\begin{itemize}
\item Virtually any  spectrophotometric standard star can be used to derive a template PSF, which then allows an empirical estimate of the PSF at the reference positions (referred to as ``\textit{nod}" positions\footnote{See the \textit{Spitzer/IRS} observer's manual at \textit{http://ssc.spitzer.caltech.edu/documents/SOM/}}). This method is currently used by the software \texttt{SPICE} (see Narron et al.\ 2007). The resulting extraction undeniably provides better S/N as compared to a tapered column extraction, although the corresponding algorithm is sensitive to cross-dispersion offsets between a source position and the nod position. A simple shift of the PSF is not sufficient to acquire the best S/N possible, and in some cases, low-frequency oscillations can appear in the spectrum due to the misalignment. The reasons are inherent to the data used to compute the PSF template since the latter is created in a specific observational mode (default positions along the slit, default data sampling). 
\item Analytical PSFs allow estimating the instrumental profile for any position along the slits. The independent effort from the ``Core to Disks" legacy program (Evans et al.\ 2003) uses such PSFs along with an extended emission background which is determined on-the-fly (see Lahuis et al.\ 2007). Analytical PSFs can bear some uncertainties due to the lack of knowledge of the exact instrumental profile.
\end{itemize}

We have developed a new optimal extraction algorithm to be used with either of the low-resolution modules of the \textit{Infrared Spectrograph} (\textit{IRS}, Houck et al.\ 2004) onboard the \textit{Spitzer Space Telescope} (Werner et al.\ 2004). The new extraction algorithm is available \textit{via} the plugin \texttt{AdOpt}\footnote{The documentation is available at \textit{http://isc.astro.cornell.edu/SmartDoc/SmartOptimal}.}, part of the new release of \texttt{SMART}. In a nutshell, an empirical super-sampled PSF has been constructed for each row of the detector array and a multi-linear regression algorithm is used to weight the pixels and derive the flux from the source. The optimal extraction is based on detector rows rather than pseudo-rectangles to treat each pixel as indivisible and thus avoid uncertainties due to the lack of knowledge on the pixel response function. The algorithm presented in this paper produces a considerably higher S/N, up to a factor of $\sim2$ for faint sources, than the current extraction method available in \texttt{SMART} for point-like sources (``tapered column"). An additional advantage of using a super-sampled PSF is that it remains valid anywhere along the aperture in the cross-dispersion direction. We found that the optimal extraction method is extremely sensitive to offsets between a source position and the nominal position. For this reason a new algorithm to locate the source in the slit has been implemented, with a precision of better than a twentieth of a pixel. It is also now possible to extract spectra of spatially blended sources, which is crucial when dealing with crowded regions, such as stellar clusters or nearby galaxies. Such as what can be achieved using iterative techniques (Lucy \& Walsh 2003), the cross-dispersion profile of the data is decomposed into its components, including the spatial profiles of any number of sources in the slit and the extended background emission. Finally, it is also possible to extract sources with significant offset in the dispersion direction by calculating a modified PSF on-the-fly. 


In the following section the basic steps to construct the PSF are given. Section\,\ref{sec:method} details the mathematical description of the method and its application for extracting multiple sources are detailed. Section\,\ref{sec:applications} briefly explains some specific applications. 

\section{Super-sampled PSF}\label{sec:psf}

\begin{table*}
\begin{center}
  \caption{Main properties of the low-resolution modules of \textit{Spitzer/IRS}.}
  \label{tab:modules}
  \begin{tabular}{c c c c c}
  \hline
  Module & Order & $\lambda$ (\mic) & Aperture size ($"$) & Pixel size ($"$) \\
  \hline
  SL & 1 & 7.4 - 14.5 & 3.7$\times$57 & 1.8 \\
  SL & 2 & 5.2 - 7.7  & 3.6$\times$57 & 1.8 \\
  SL & 3 & 7.3 - 8.7 &  3.6$\times$57 & 1.8 \\
  \hline
  LL & 1 & 19.5 - 38.0 & 10.7$\times$168 & 5.1 \\
  LL & 2  & 14.0 - 21.3 & 10.5$\times$168 & 5.1 \\
  LL & 3  & 19.4 - 21.7 & 10.5$\times$168 & 5.1 \\
  \hline
  \end{tabular}
\end{center}
\end{table*}

We have constructed super-sampled PSFs for the two low-resolution modules (Short-Low, SL and Long-Low, LL, see Table\,\ref{tab:modules}) of the IRS on \textit{Spitzer}. The optimal extraction for the high-resolution modules will be considered in the future since these modules do not allow the full sampling of the PSF profile. 

In order to build the super-sampled PSF, we considered observations of calibration stars scanned around reference positions in the cross-dispersion direction. We used the basic-calibrated data (BCD) product. After cleaning the bad pixels using \textit{IRSCLEAN}\footnote{\textit{IRSCLEAN} can be found at \textit{http://ssc.spitzer.caltech.edu}}, individual exposures (DCEs) were combined, after verifying that the pointing remained stable. The source in each image was then found \textit{via} the source finder algorithm (Sect.\,\ref{sec:sf}) which makes use of previous iterations of the PSF profile. Finally, the positions of the source along the slit were used as inputs in an iterative reconstruction of the high-resolution spatial profile from the under-sampled data. We used a simplified version of the regularized image reconstruction explained in detail by Pinheiro da Silva et al.\ (2006). In short, considering a series of $k$-undersampled spectra $D_k$ and the desired high-resolution profile $P$, the observed spectra take the form:
\begin{equation}
D_k = U W_k P + \eta_k,
\end{equation}
where $U$ is the downsampling matrix, $W$ is the geometric transformation matrix, i.e., in our case the relative shift with respect to a reference position, and $\eta$ represents the noise. $W$ is easily calculated from the source positions, while $U$ is the matrix corresponding to a simple re-scale with no interpolation. We used the same algorithm described by Pinheiro da Silva et al.\ (2006), with the following iterative step:
\begin{equation}
\hat{P_{j+1}} = \hat{P_j} + \mu \sum_k (W_k^T U^T D_k) - (W_k^T U^T U W_k) \hat{P_j} 
 - \lambda (\hat{P_j}-P_0),
\end{equation}
where $\mu$ controls the convergence speed, $\lambda$ controls the regularization, and $P_0$ is a reference spectrum, which can be either the theoretical PSF, or an initial guess based on a simple co-addition of the aligned data images.

\begin{figure*}
\epsscale{2.3}
\plottwo{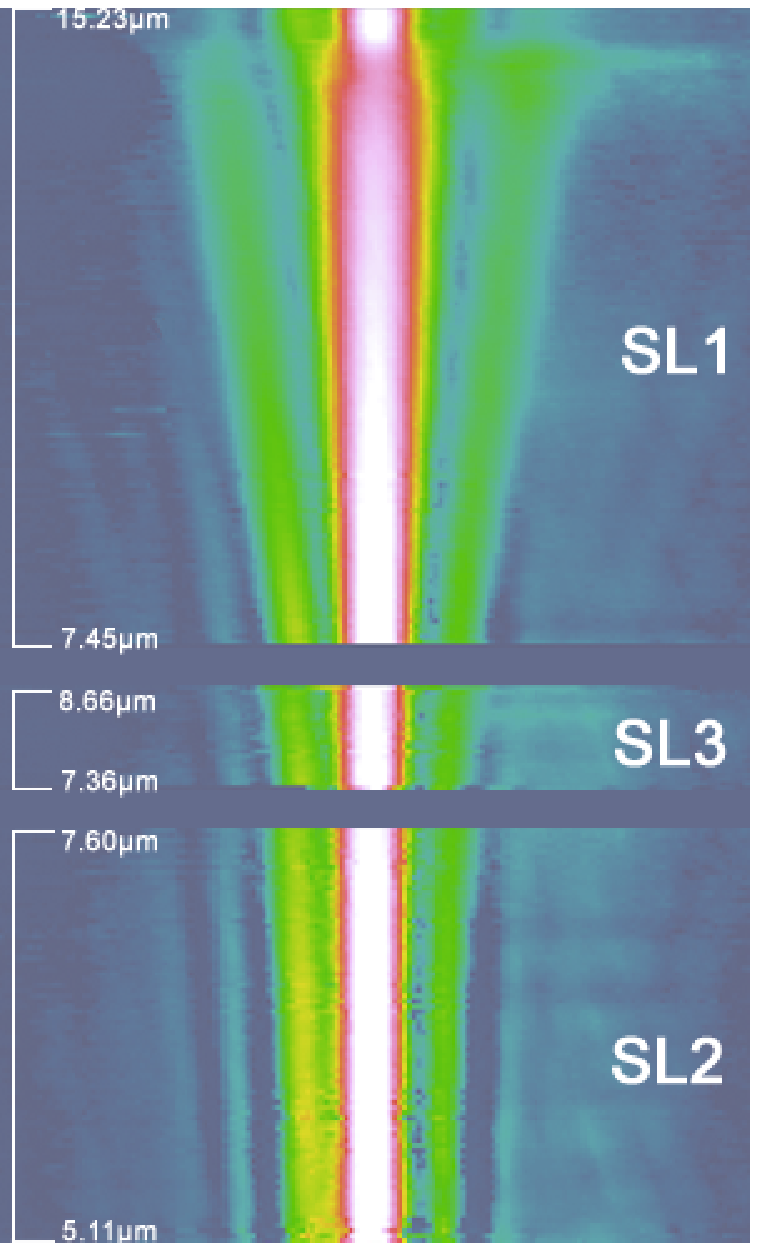}{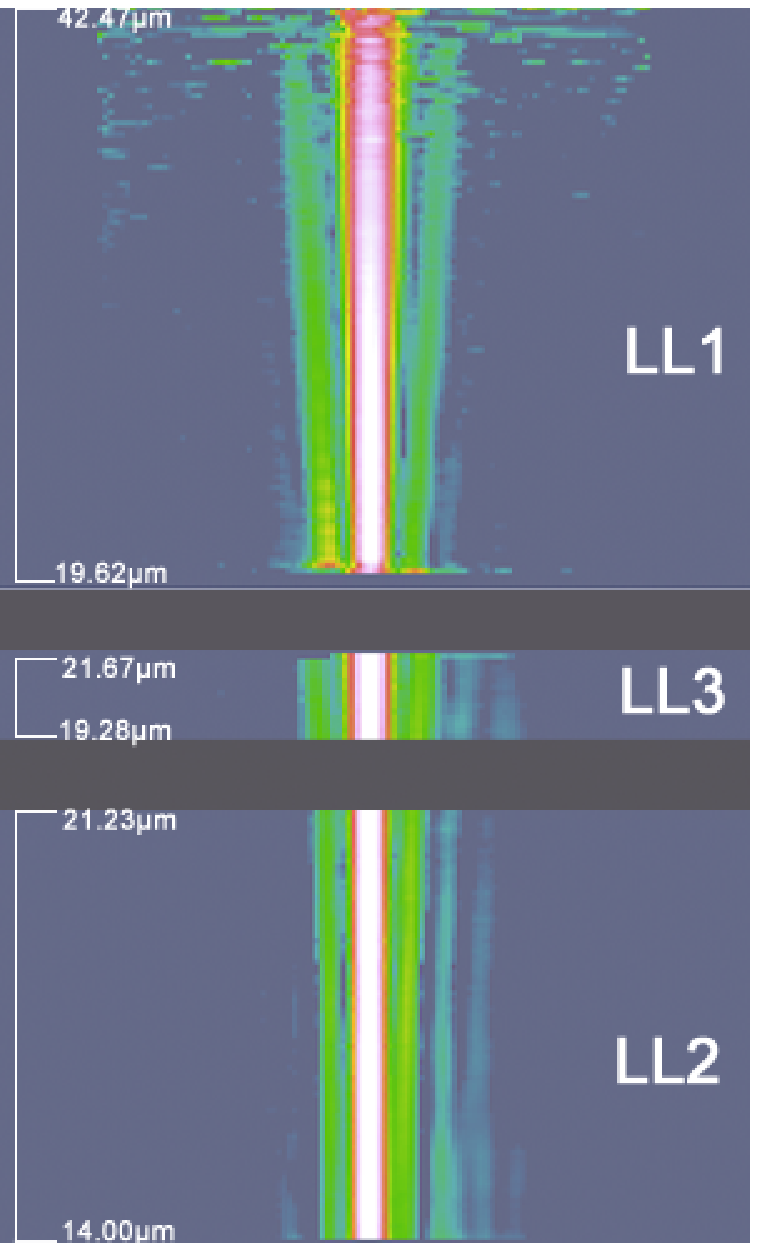}
\caption{SL and LL super-sampled PSFs.\label{fig:psfimage}}
\end{figure*}

\begin{figure*}
\epsscale{2.0}
\plottwo{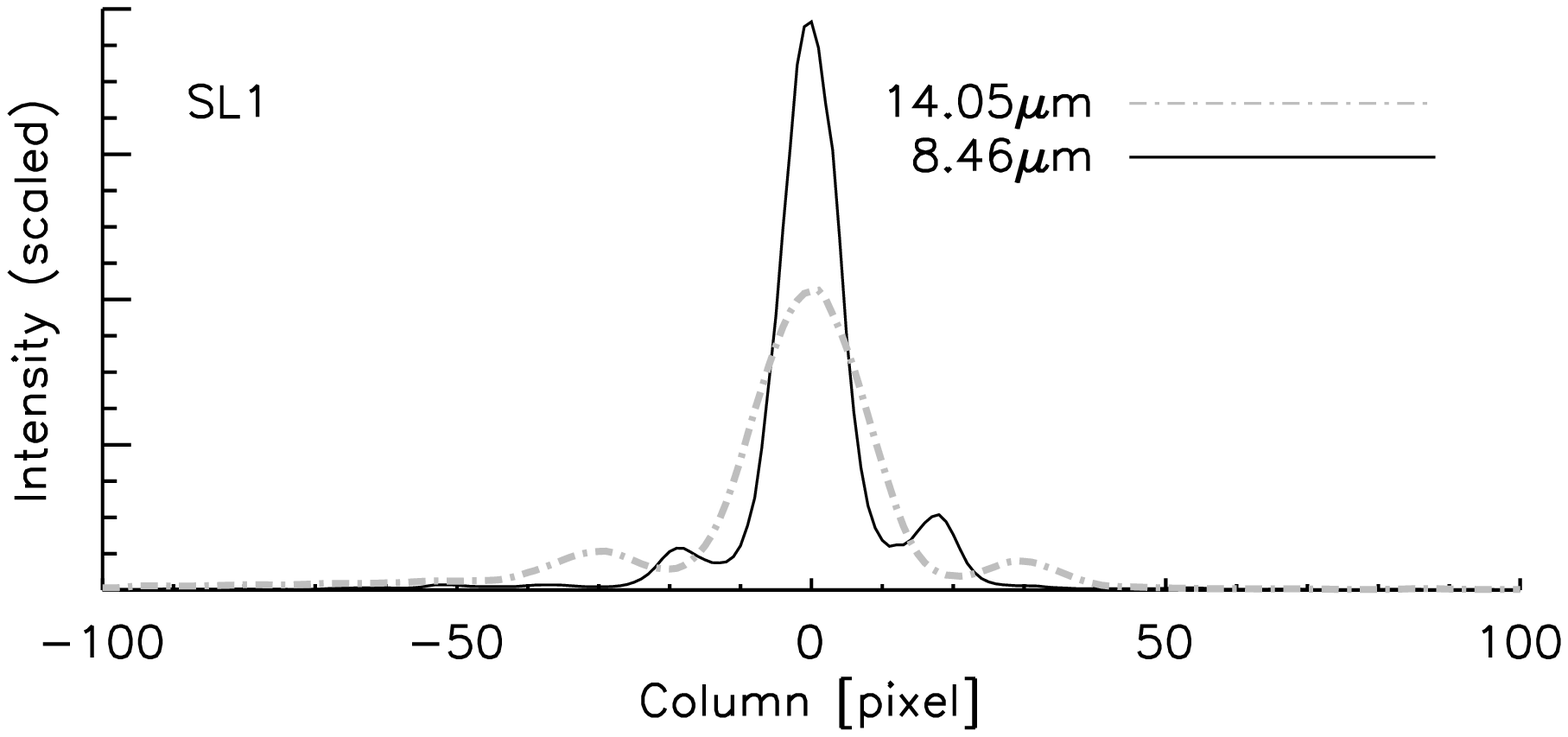}{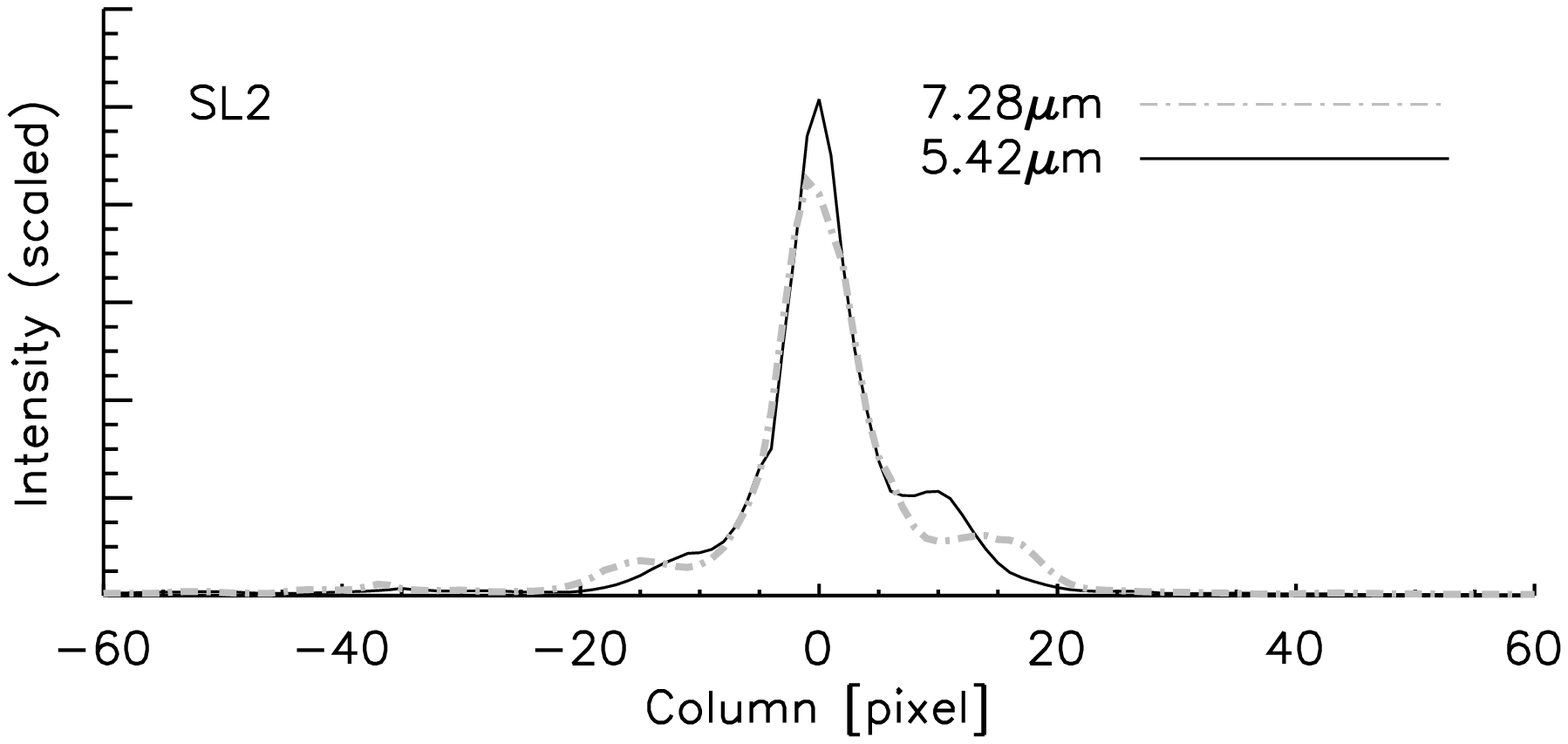}\\
\plottwo{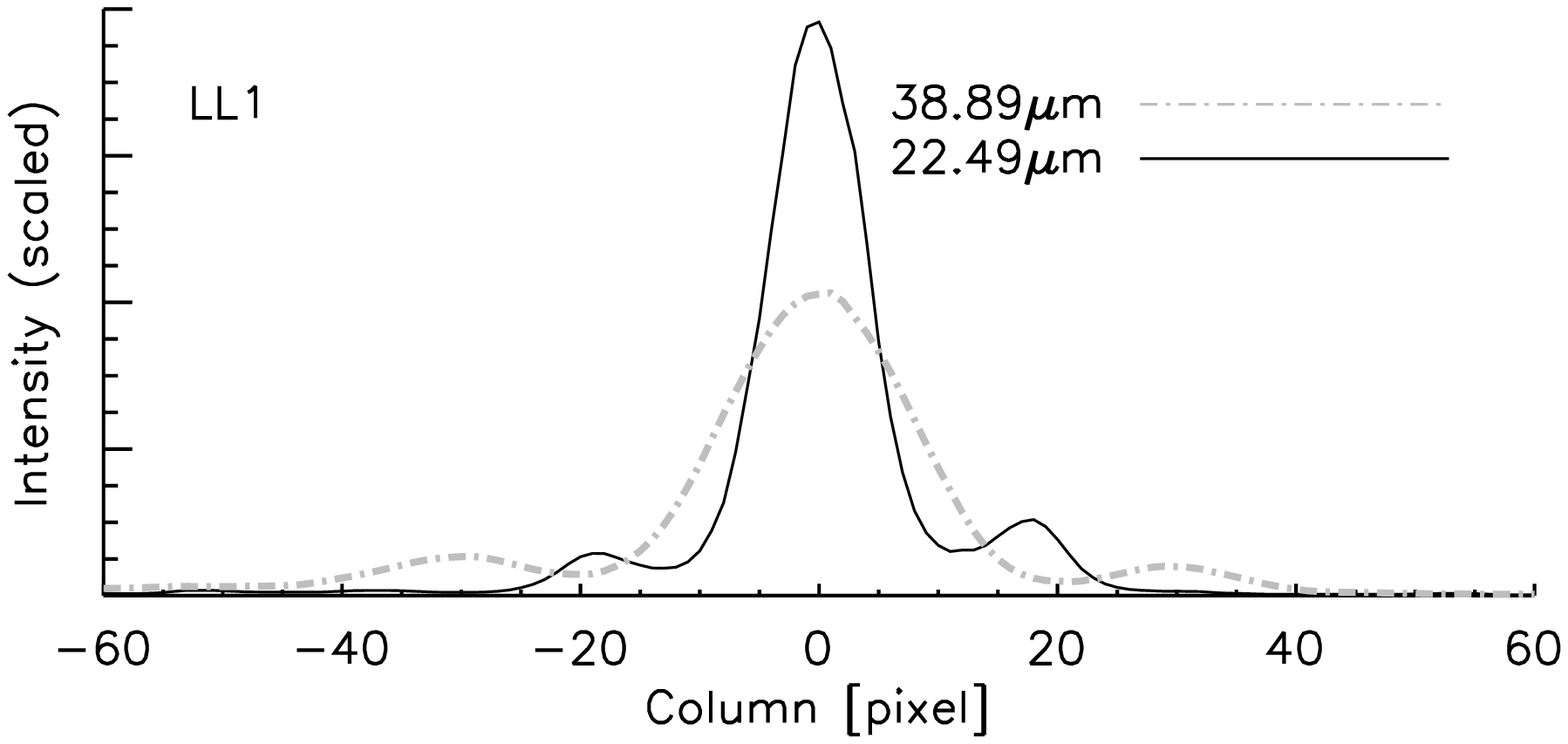}{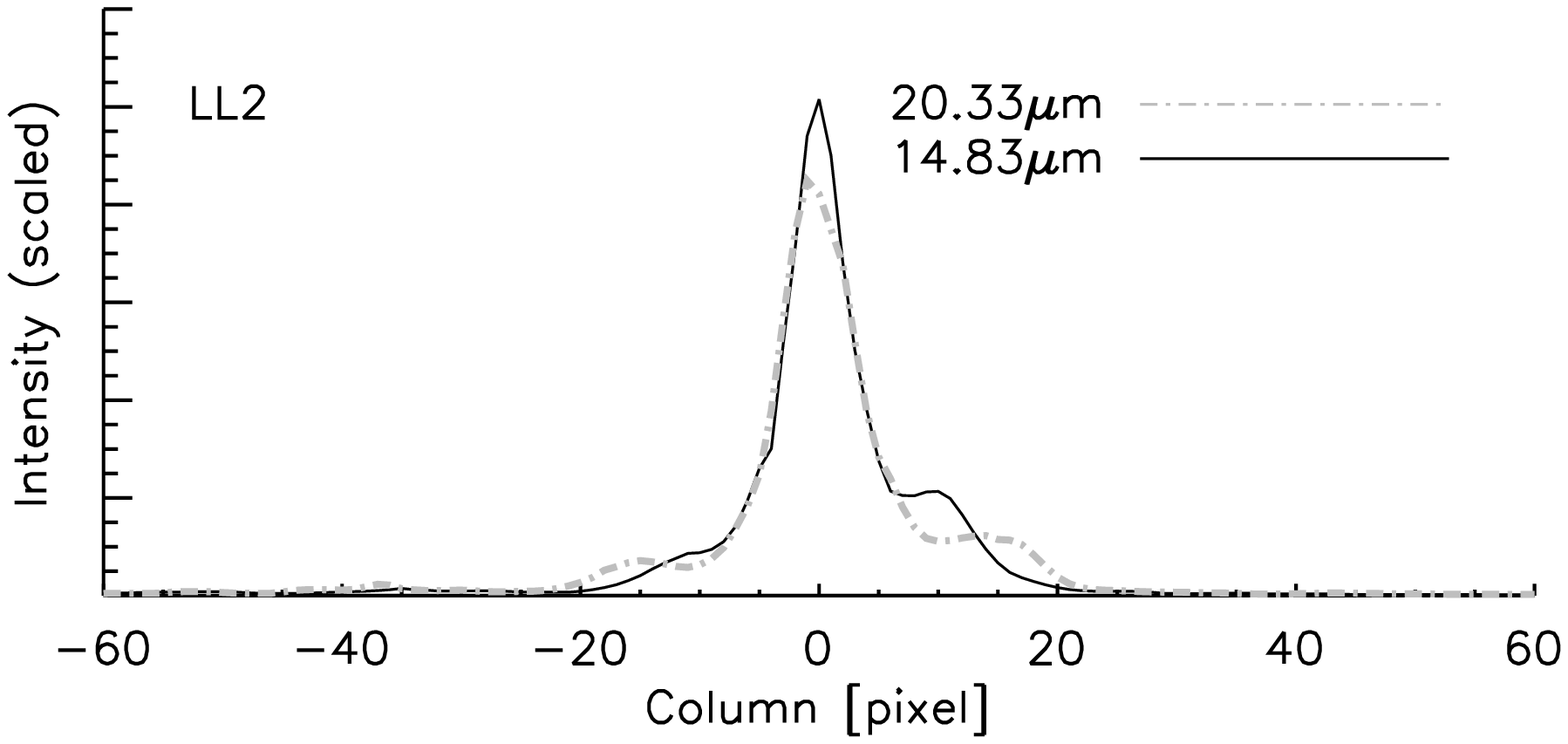}
\caption{Cut of the SL and LL super-sampled PSF along various wavelength rows.\label{fig:psfcut}}
\end{figure*}

In practice, we calculated the SL PSF on a reference grid with 10 sub-pixels per pixel per actual pixel, while the LL grid contains five sub-pixels. The number of sub-pixels was chosen to maximize the signal in each sub-pixel while allowing the best resolution possible on the PSF profile. The shape of the PSF is the same within the uncertainties for the two nod positions. Table~\ref{tab:nods} gives the coordinates for the {\em nod} positions and the equations defining the spectral trace in each module. The spectral trace quantifies how the position of the centroid of the PSF shifts with wavelength (i.e., up and down the array). Its shape differs slightly in the two nod positions because the nod positions are at different distances from the center of the focal plane. The extraction algorithm interpolates or extrapolates the trace polynomial coefficients depending on the source position. Figure\,\ref{fig:psfimage} presents the PSF for each module. Our reconstruction of the PSF is good enough to reveal multiple Airy rings. Figure\,\ref{fig:psfcut} displays cuts along various wavelength rows.

\begin{deluxetable}{c c c}
\tablewidth{0pc}
\tabletypesize{\scriptsize}
\tablecaption{Nod position coordinates and trace equations.\label{tab:nods}}
\tablehead{
\colhead{Field of view} & \colhead{Nod position} & \colhead{Equation}  }
\startdata
  SL1 nod 1		& 9.26	& $(4.51\times10^{-2})  y  - (2.21\times10^{-5})  y^2 $\\
  SL1 nod 2		&19.85	& $(4.51\times10^{-2})  y  - (2.21\times10^{-5})  y^2 $\\
  \hline
  SL2 nod 1		& 55.53	& $(4.32\times10^{-2})  y  - (3.71\times10^{-7})  y^2 $\\
  SL2 nod 2		& 65.52  & $(4.32\times10^{-2})  y  - (3.71\times10^{-7})  y^2 $\\
  \hline
  SL3 nod 1		& 52.34 & $(4.50\times10^{-2})  y  - (1.90\times10^{-5})  y^2 $\\
  SL3 nod 2		& 62.46  & $(4.50\times10^{-2})  y  - (1.90\times10^{-5})  y^2 $\\
  \hline                                                 
  \hline    
  LL1 nod 1		& 36.29 & $(-5.10\times10^{-2})  y  + (4.25\times10^{-5})  y^2 $ \\
  LL1 nod 2		&  47.00  & $(-4.71\times10^{-2})  y  + (3.93\times10^{-5})  y^2 $ \\
  \hline
  LL2 nod 1		& 74.02	  & $(-6.69\times10^{-2})  y  - (5.58\times10^{-5})  y^2 $ \\
  LL2 nod 2		& 85.00  & $(-6.16\times10^{-2})  y  - (5.14\times10^{-5})  y^2 $\\
  \hline
  LL3 nod 1		& 74.09	& $(-6.69\times10^{-2})  y  - (5.58\times10^{-5})  y^2 $  \\
  LL3 nod 2		& 85.20  & $(-6.16\times10^{-2})  y  - (5.14\times10^{-5})  y^2 $\\
\enddata
\tablecomments{Nod positions are given in pixels and are derived from the trace intersect at row 0. $y$ represents the row number in the detector array.}
\end{deluxetable}

\section{Methodology}\label{sec:method}

\subsection{Wavelength grid}

When performing the spectral extraction, the algorithm considers each row independently. The wavelength actually varies along a given row so that a different location in the spatial direction corresponds to a slightly different wavelength. In theory the use of pseudo-rectangles is adequate to consider windows with uniform wavelength but the practical implementation is limited by the lack of precise knowledge of the response function within a pixel (see introduction). Furthermore, a slight loss of spectral resolution occurs due to the artificial splitting of information in individual pixels between wavelength elements. Regardless of these side-effects, considering rows instead of pseudo-rectangles provides a valuable means to better sample the resulting spectrum by using the exact wavelength at the nod positions.

\begin{figure}[b!]
\epsscale{1.0}
\plotone{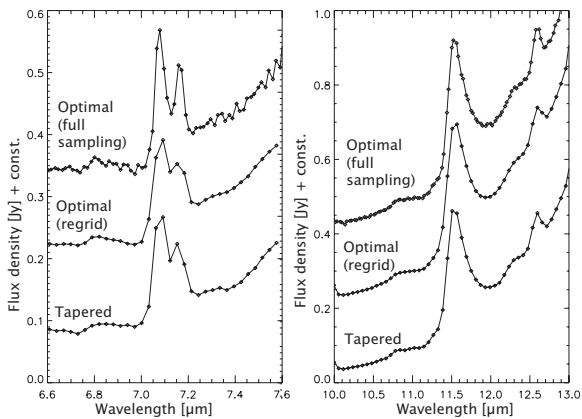}
\caption{SL spectrum of the galaxy NGC\,6240 resulting from the combination of the two nod spectra. The blended lines H$_2$ and [Ar\2] are shown on the left while the PAH feature at 11.3\mic\ is shown on the right. }
  \label{fig:ngc6240}
\end{figure}

Figure\,\ref{fig:ngc6240} shows an example of a spectrum resulting from the combination of two nod spectra in which the wavelength is determined directly at the source location in the cross-dispersion direction. The slight wavelength shift between the two nod positions provides a better spectral sampling at the expense of a lower S/N ratio because the nod spectra are not combined in the same wavelength scale. Narrow spectral lines benefit greatly from the full sampling as their profiles are better determined, implying a better detection level, and a more accurate line flux measurement. It is necessary that the two nod spectra align fairly well for them to be combined. For this reason, we implemented an automatic algorithm which, provided two nod spectra, fits the error function (difference between the spectra) with a $b$-spline. The interpolated difference is then split evenly to the individual spectra which can then be combined. This algorithm preserves the spectral resolution. Practically, the \texttt{SMART} software can regrid \textit{a posteriori} the data on the reference wavelength grid or keep instead the wavelength determination at the source position (see online documentation).

\subsection{Optimal extraction}\label{sec:optimal}

Before the PSF can be used to perform the optimal extraction, it has to be downsampled and shifted to the source position. This important process consists of three steps: (1) the data grid (sub-array of the original 128-pixel row where the spectral order is the one requested) is stretched to match the (super-sampled) PSF grid and is shifted to align the PSF and the source. (2) The PSF is interpolated in the new grid. (3) The PSF flux of a given pixel in the data grid is determined by integrating the flux of the subpixels.

To find the flux from the source in each row, we use a multiple linear regression algorithm. The scaling factor $f(\lambda)$ gives the flux for a given row, and is thus constant along the row. Each pixel can thus be used to determine $f(\lambda)$ independently:
\begin{equation}
f(\lambda) = D_i / P_i\ \forall i\in[0, n],
\end{equation}
where $i$ is the column element of the slit spatial profile,  $D$ is the source profile, and $P$ is the PSF profile. The scaling factor $f(\lambda)$ provides the source flux in e$^-$/sec, which eventually leads to the source flux density in Jy after calibration. The flux calibration performed by \texttt{AdOpt} uses the calibration files \texttt{b\{0 2\}$\_$aploss$\_$fluxcon.tbl} provided by the \textit{SSC}. These files provide a calibration which does not include extraction aperture-loss light corrections due to the size of the regular extraction window. It is thus well suited for the purpose of optimal extraction which does not involve a regular aperture. An additional relative spectral response function (RSRF) is used to correct for the small ($\sim$10\%) residual flux offsets. The RSRF used in \texttt{SMART} is calculated by performing optimal extraction on the calibrator star HR\,7341 which is then compared to its stellar theoretical template provided by the \textit{SSC}\footnote{\textit{http://ssc.spitzer.caltech.edu/IRS/calib/templ/}} (see also Decin et al.\ 2004). The flux calibration may be subject to slight modifications in the future as new versions of calibration files will be provided by the \textit{SSC}.

\begin{figure*}
\includegraphics[angle=0,scale=0.95]{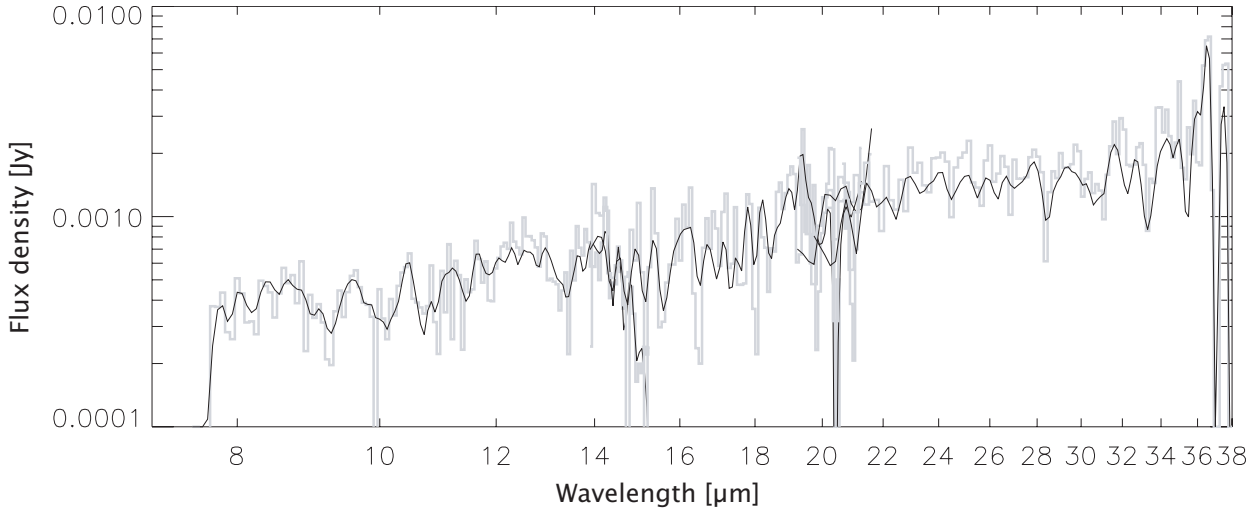}
  \caption{Illustration of the S/N improvement provided by optimal extraction. The object is the galaxy NOAO14213.5+35.1. The histogram shows the tapered column extraction while the solid line spectrum shows the optimal extraction. The improvement in the S/N ratio is a factor $\sim1.8$ in this case.}
  \label{fig:sn}
\end{figure*}

The use of optimal extraction increases the overall quality of the spectra, with the most significant improvements for observations with low S/N. Quantitatively, the S/N ratio increases by a factor $\sim1.5-2.0$ for sources as faint as 1\,mJy compared to normal tapered column extraction. In addition, while tapered column extraction is sensitive to bad pixels, optimal extraction is mostly unaffected, except when  the bad pixels are precisely located at the peak of the spatial flux distribution.

\subsection{Multi-source extraction}\label{sec:multi}

\begin{figure}
\includegraphics[angle=0,scale=.32]{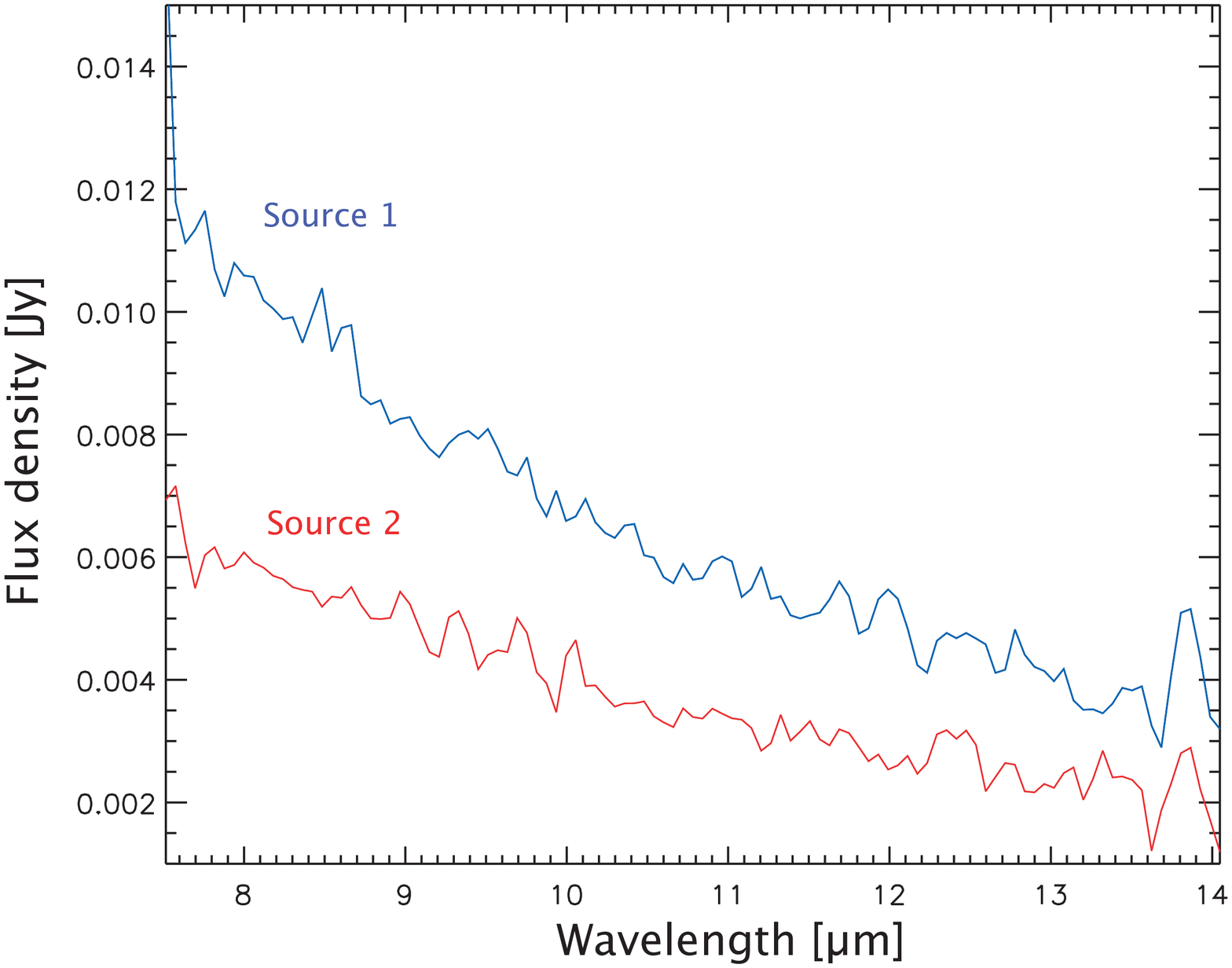}
\includegraphics[angle=0,scale=.42]{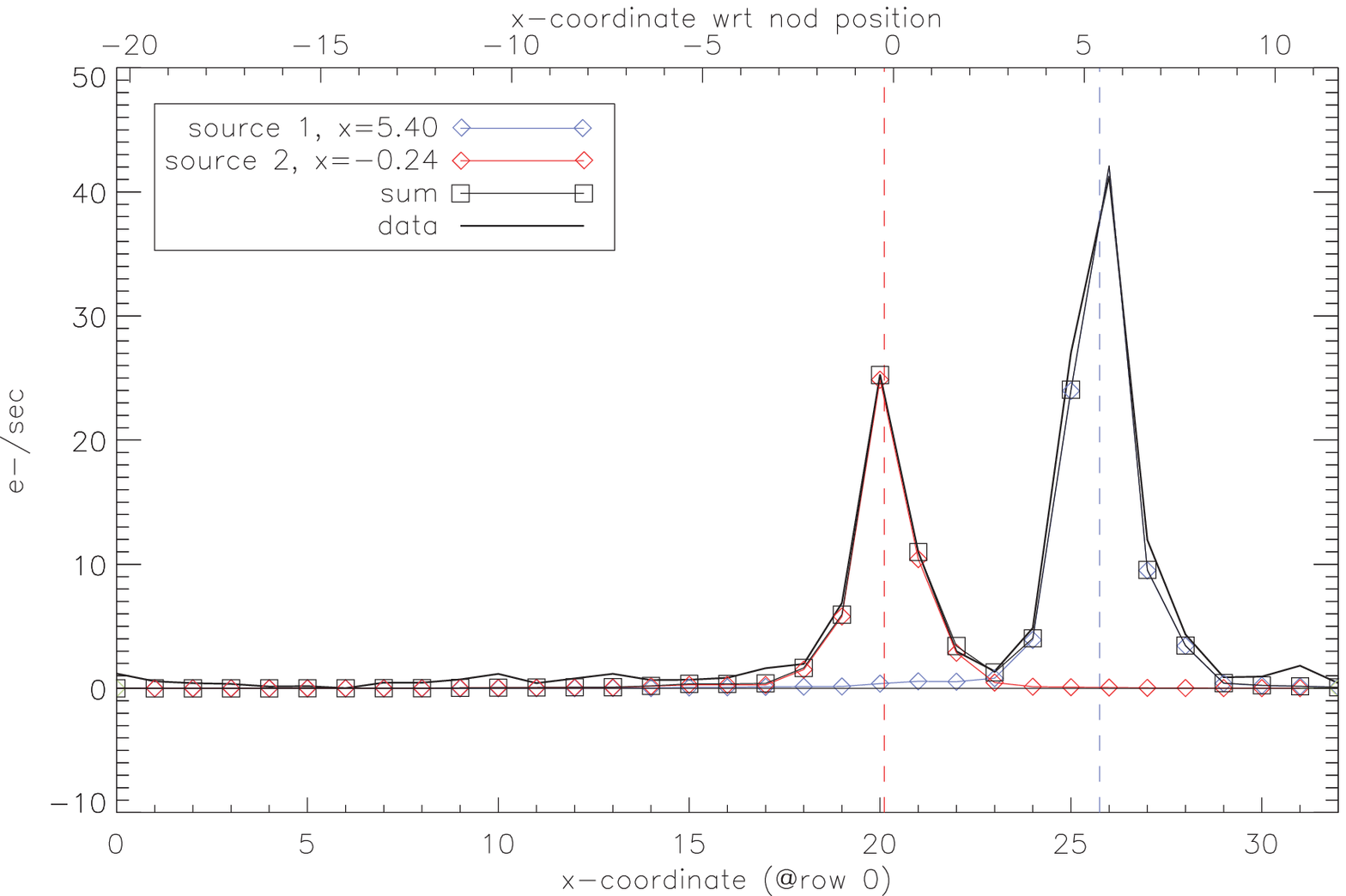}
  \caption{Real-case scenario of slightly blended sources in the Small Magellanic Cloud. The spatial profile is calculated based on the collapse of the shortest wavelengths, i.e., where the PSF is the narrowest.}
  \label{fig:ms1}
\end{figure}

\begin{figure}
\includegraphics[angle=0,scale=.32]{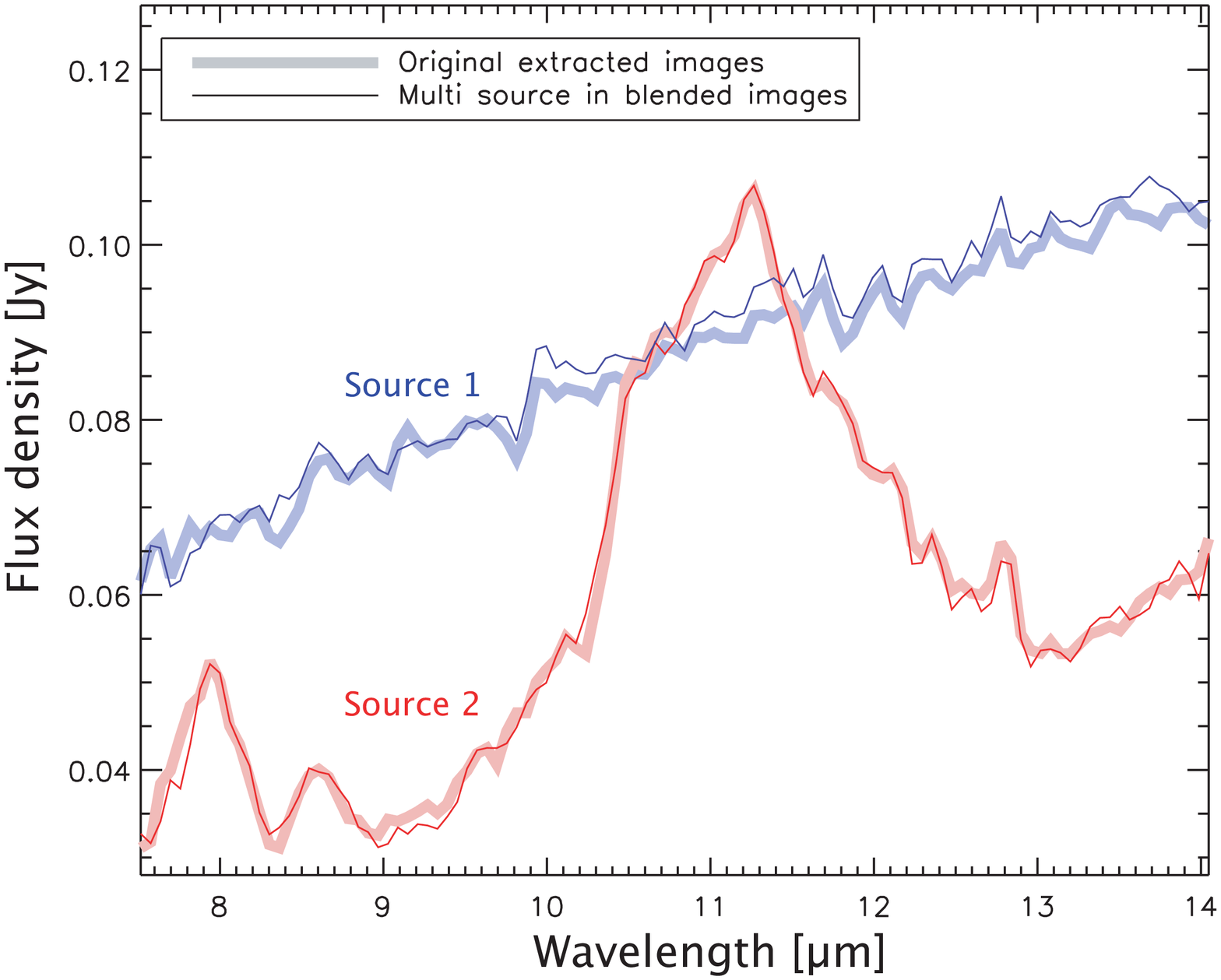}
\includegraphics[angle=0,scale=.42]{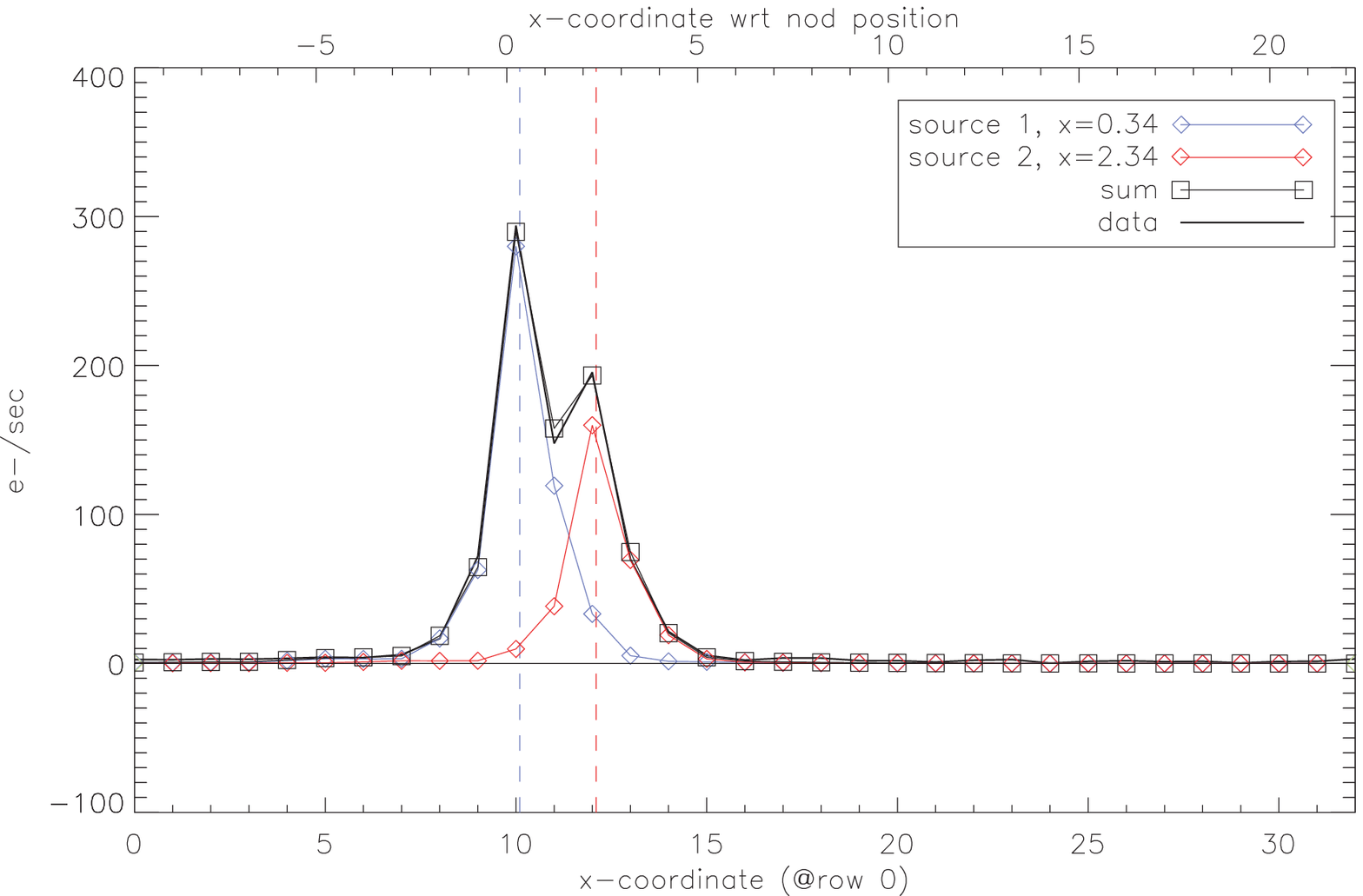}
  \caption{Spectra extracted from an artificial image in which spectral images of the quasar 3C371 and the planetary nebula SMP-LMC35 have been combined after shifting one arbitrarily by two pixels in the cross-dispersion direction. The resulting spatial profile is shown on the right. }
  \label{fig:ms2}
\end{figure}

One of the most interesting applications enabled by the knowledge of the super-sampled PSF is the possibility of extracting spatially-blended sources. In the case of multiple sources, the estimated spatial profile is given by
\begin{equation}
\hat{D_i} = \sum_k^m f_k(\lambda) \times P_i^{(k)},
\end{equation}
where $m$ is the number of sources, $f_k$ is the fit parameter, and $P^{(k)}$ is the PSF aligned with the $k$-th spectrum. As explained by Collins, Gull, Bowers \& Lindler (2002) and Bevington \& Robinson (1992), the coefficients $f_k$ can be found by solving the multiple linear regression using the least-square method:
\begin{equation}
\hat{f} = (P P^T)^{-1} P^T D,
\end{equation}
where $D$ is the vector $\{D_i\}$ and $P$ is the matrix $\{P_i^{(k)}\}$. 
The system is characterized by $n$ equations (number of columns where the order is as requested) for $m$ unknowns. The system thus remains overdetermined for a large number of sources. We find that it is possible to extract spectra of sources separated by two pixels or more along the spatial axis. Sources separated by one pixel can be extracted, with recognizable spectral features, but the measurement of the features (integrated flux, equivalent width) should be regarded with caution. As a result of the minimum separation between two source and of the system overdetermination, the algorithm can extract a number of sources which is less than half the number of pixels in the row (minus the number of degrees of freedom for the background, see Sect.\,\ref{sec:bg}). Thus, theoretically the maximum number of sources that can be extracted is $\approx17$ for SL to $\approx15$ for LL. Figure\,\ref{fig:ms1} gives an example of a real-case scenario in the Small Magellanic Cloud, while Figure\,\ref{fig:ms2} represents an artificial case for which the recovered spectra are similar to the original spectra used to construct the artificial test image.

There are generally two main scenarios adapted to source deblending. 
\begin{itemize}
\item[1.] Sources can be known \textit{a priori} in which case their position on the sky can be used to extract their spectra. Sources can be also identified in the detector image when spatial structure is evident. Note however that the multi-source extraction is originally meant only for point-like sources. 
\item[2.] Sources can be discovered with the help of the source finder (residual minimization, see Sect.\,\ref{sec:sf}).
\end{itemize}

Because the multi-source extraction is an interesting tool for analysis of crowded regions, the \texttt{AdOpt} software allows the user to display the slit projected on the sky along with the field of view position and the requested coordinates. \texttt{Simbad}\footnote{\texttt{Simbad} is found at \textit{http://simbad.u-strasbg.fr/simbad/}} and \texttt{NED}\footnote{\texttt{NED} is found at \textit{http://nedwww.ipac.caltech.edu/}} sources in the field can also be displayed for easy source identification. Finally, any FITS image can be overplotted, including automatic queries from the SDSS, 2MASS, Spitzer/IRAC, and Spitzer/MIPS catalogs.

\subsection{Spectral-pointing induced throughput error (SPITE)}\label{sec:spite}

No calibration currently exists for offsets in the dispersion direction. Such offsets can result in significant light loss in the SL module when sources are more than $\sim1"$ from the slit center (see \textit{IRS} reports by Sloan 2004; Sloan, Nerenberg \& Russell 2003; Nerenberg \& Sloan 2003). This effect is known as the spectral-pointing induced throughput error (SPITE). Since the LL aperture is much wider than the SL one (Table \ref{tab:modules}), the effect is unimportant for LL except for significant offsets. 

Accounting for offsets in the dispersion direction keeps the optimal extraction valid for sources located anywhere in the aperture. It provides a reliable flux calibration which is useful in the case of a mispointing, and it is also adapted to the case of multiple sources as it is likely that at least one source is not near the slit center. While a regular tapered extraction would require only the knowledge of the flux fraction that is lost outside the slit (provided by the \texttt{b\{0 2\}$\_$slitloss$\_$convert.tbl} tables from the \textit{SSC}), optimal extraction requires the knowledge of the PSF profile at the exact source position. Available data could not be used to derive the PSF profile at various positions in the dispersion direction. We used instead a theoretical PSF and calculated the function transforming the ``nominal PSF" (i.e., integrated PSF across the slit for a source at the slit center) to the effective PSF (source anywhere across the slit). The function is then applied to the empirical super-sampled PSF derived (Sect\,\ref{sec:psf}). Finally, the effective PSF is normalized to the total area of the nominal PSF so that the scaling factor provides the correct flux value without any further calibration required other than the one already applied.

  \begin{figure}
\includegraphics[angle=0,scale=.5]{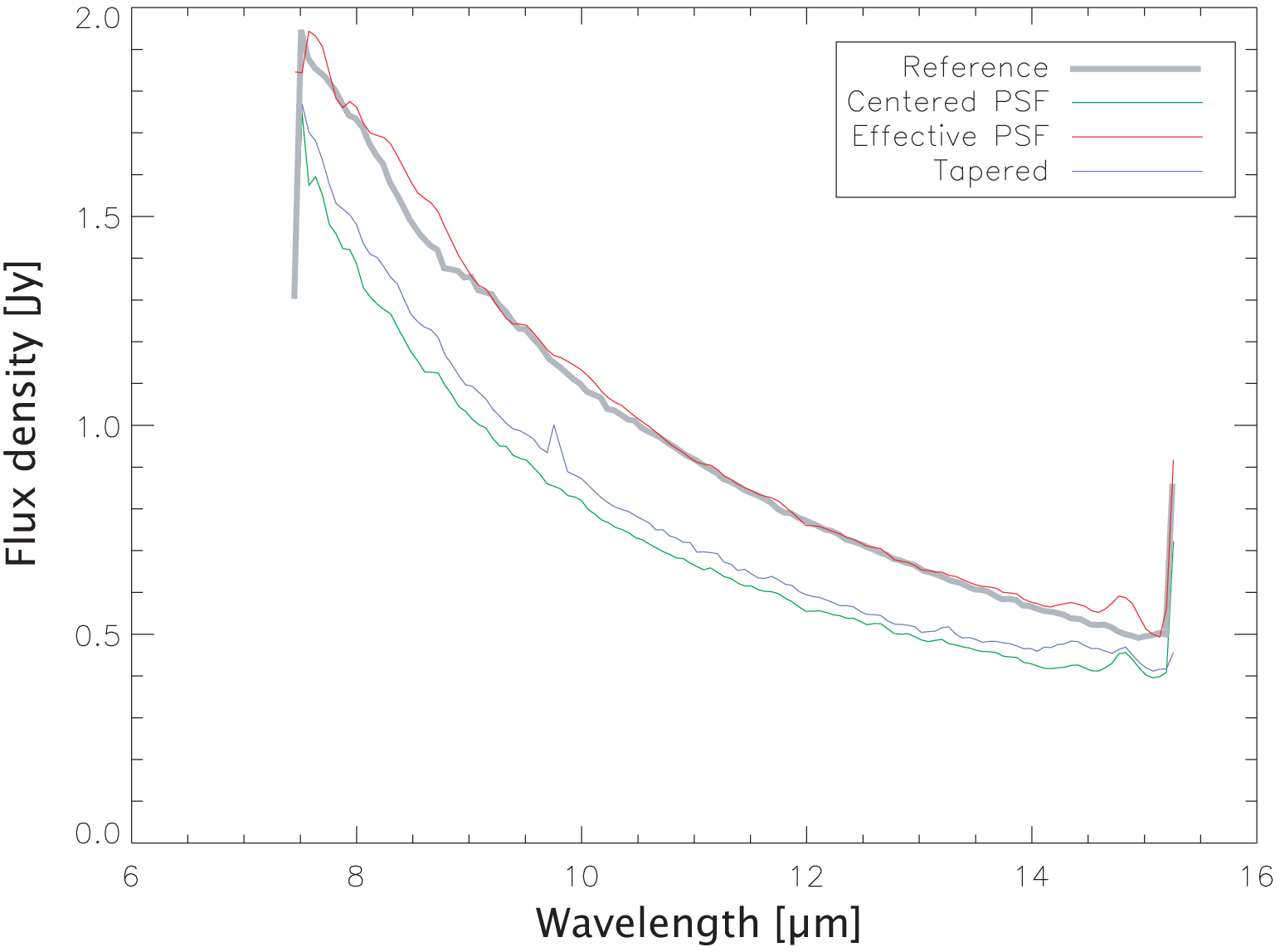}
\includegraphics[angle=0,scale=2.3]{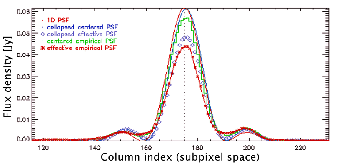}
  \caption{The SL spectrum of the star HR\,6348 is shown on the left. The reference spectrum shows the median spectrum of all the existing observations of this star. The ``Centered PSF" spectrum shows the extraction of the mispointed exposure assuming that the PSF is centered in the dispersion direction. The ``Effective PSF" shows the extraction of the same image, by calculating the PSF on-the-fly at the real source position. The corresponding PSF profiles are shown on the right. }
  \label{fig:hr6348}
\end{figure}

With the throughput error corrected, it becomes possible to extract sources anywhere in the slit with a reliable flux calibration. \texttt{SMART} allows the user to extract sources at specific celestial coordinates, but can also find the source in the dispersion direction. This is possible because as the source moves away from the slit center, the effective PSF becomes wider so that the quality of the fit varies. Figure\,\ref{fig:hr6348} shows an example on an actual mispointing of $\approx$1.6$"$ in the dispersion direction of the calibration star HR\,6348. The offset is about half the SL aperture height, providing an adequate illustration of the algorithm explained above. By using an on-the-fly PSF, the flux level agrees with the true spectrum while providing smaller residuals (by 20\%).

\subsection{Complex background}\label{sec:bg}

There are usually several ways to remove the background. The background contribution can be inferred by interpolating the columns on each side of the source, which requires good knowledge of the source position. Another way to remove the sky relies on the subtraction of dedicated background observations or differencing the image by {\em nod} or by order (e.g. SL nod 1 minus SL nod 2, or SL1 nod 1 minus SL2 nod 1). The disadvantage of these methods is that they assume that the background at the position where it is measured is the same to where the source is located. The new algorithm can calculate the underlying extended emission together with the PSF scaling factor (see also Geers et al.\ 2006, Lahuis 2007). Therefore, even the pixels where the source emits are taken into account for constraining the background shape and level.

\begin{figure}
\epsscale{1.0}
\plotone{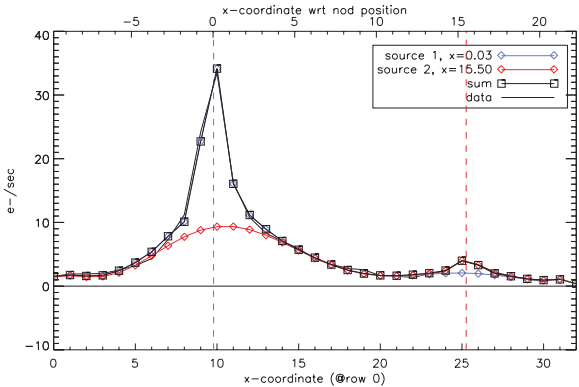}
  \caption{Real-case example of how a complex background is handled by \texttt{SMART}. The background is determined together with the optimal extraction of the sources (two stars in the Small Magellanic Cloud).}
  \label{fig:bg}
\end{figure}

The \texttt{AdOpt} algorithm assumes that every column in a given spatial row should follow the same background equation. Any background shape can be considered which is a function of the column index $i$, i.e., with $B_i = \sum_l \alpha_l \times g(i, l)$ with the coefficients $\alpha_l$ being the unknowns and $g(i, l)$ represents any function that describes the background. We thus have again $n$ linear equations:
\begin{equation}
\hat{D_i} = \sum_k^m f_k(\lambda) \times P^{(k)}_i + \sum_l \alpha_l \times g(i, l).
\end{equation} 
The system is solved using the multi-linear regression algorithm, which estimates simultaneously the unknowns $f_k$ and $\alpha_l$. For practical reasons, the algorithm \texttt{AdOpt} makes use of a polynomial background, i.e., $g(i, l) = i^l$. The system remains overdetermined if the sum of the number of sources and the background parameters does not exceed $\approx17$ for SL to $\approx15$ for LL. An example of an optimal extraction of a source within a complex background is shown in Fig.\,\ref{fig:bg}.

The particularity of this method is that the background profile is estimated for each row independently, which can lead to undesirable small-scale variations along the wavelength axis. Hence for noisy images, it is advised to perform a first extraction to estimate the row-by-row background and then smooth the reconstructed background image along the wavelength axis (see online documentation). 

  \subsection{Source finder}\label{sec:sf}

\begin{figure}
    \includegraphics[angle=0,scale=0.33]{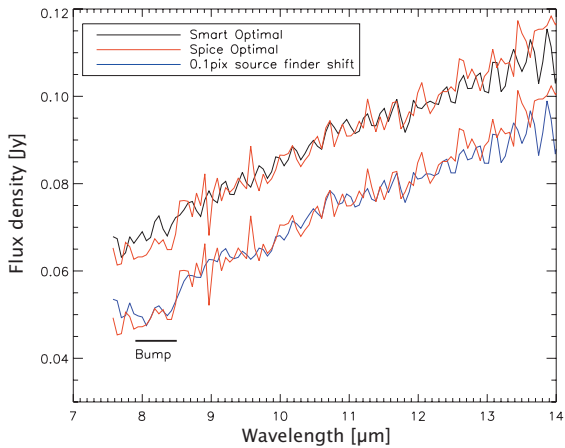}
  \caption{Spectrum of the quasar 3CQ-3C371 resulting from the extraction of a single DCE image. \textit{Top} $-$ The dip at short wavelengths observed is typical of the artifacts introduced by slight uncertainties in the source position coupled with a normally sampled PSF. \textit{Bottom} $-$ The artifact could be exactly reproduced by using the super-sampled PSF of \texttt{SMART} and a forced shift of 0.1 pixel.}
  \label{fig:bump}
\end{figure}

  The plug-in \texttt{AdOpt} includes a new source-finding algorithm. Several source finders already exist, using for instance the centroid in the collapsed spatial profile (along the PSF trace),
  or the median of the centroids in each row. These methods require fitting a profile, usually using a Gaussian curve, which inevitably introduces possible systematic errors and even biases. These biases can introduce negative effects in the data, such as noticeable bumps in the spectrum (which are more abrupt the larger the discrepancy between the source finder and the source actual position) (see example in Figure\,\ref{fig:bump}).

 The new source finder uses the super-sampled PSF. The PSF is aligned at varying positions along the slit. For each position, the residual image (data image minus the PSF image) is then analyzed. First, the residual image is rectified using the trace equations of Table\,\ref{tab:nods}. Then, a $b$-spline is fitted to each column, with rejection parameters such that bad pixels are identified and ignored. The median value of the fit is then calculated for each column, which results in a representative spatial row. This row is ideally zero when there is no background and when the PSF has been aligned to the source. Hence, by simply taking the average of this row, one has a very good leverage on the quality of the alignment PSF-source.  The precision reached by the source finder is better than $1/20^{\rm th}$ of a pixel.

\section{Other applications}\label{sec:applications}

Besides the improved S/N and the multi-source extraction, the precise knowledge of the PSF profile enables several valuable possibilities for the handling of point-like sources while providing interesting tools for handling extended sources.

\subsection{Spatial extent}

With the knowledge of the super-sampled PSF, the spatial extent of a given source can be determined fairly accurately by comparing its spatial profile to the PSF for the same wavelength or wavelength range. The plugin \texttt{AdOpt} allows the user to check the extent of any source within the slit, by providing three meaningful quantities: the percentage of the extent with respect to the PSF, the extent in pixels, and the extent in arcseconds. Among the most useful applications to check the source extent is to identify the possible presence of another source or the presence of extended emission associated with the point-like source (see \ref{sec:pls_ext}).

\subsection{Point-like sources embedded within an extended source}\label{sec:pls_ext}

The complex case of point-like sources embedded within an extended emission can be approached in various ways depending on the background nature.
\begin{itemize}
\item \textit{Large-scale background with spectral features}. A low-grade polynomial background can be estimated in every wavelength row (Sect.\,\ref{sec:bg}). The image quality should however be better than some threshold.
\item \textit{Large-scale background with no spectral features}. The same procedure as above is applied, but the final background image is calculated by smoothing the ``row-by-row" background. 
\item \textit{Small-scale background (e.g., disk, extended region, ...)}. A polynomial of high degree should be used, implying that less constraints will be possible. In fact, a polynomial might not suffice to adequatly represent the background, but it will still allow a first order approximation (see next section).
\end{itemize}

\subsection{Removal of point-like source contribution}

In the case of a point-like source embedded in an extended source, it is possible to extract the spectrum of the extended source alone by extracting first the point-like source (no background subtraction in this case). The reconstructed image (which in the
  case of images with normal point sources contains the residuals) includes only the extended source.  Note that the spectrum of the point-like source includes a fraction of the extended emission and should be
  regarded with caution, especially for flux calibration and spectral features measurements. The extended emission in the reconstructed image can then be extracted using either a full slit or a tapered column.  It is important to bear in mind that there is no flux calibration available for the extended emission, except if the source fills uniformly the aperture.

\section{Summary}

We have presented the optimal extraction algorithm adapted to the \textit{Infrared Spectrograph} onboard \textit{Spitzer}. Besides providing a significant increase in the signal-to-noise ratio as compared to the regular ``tapered" extraction, the new algorithm allows extraction of multiple sources at any location within the aperture. The optimal extraction is also implemented for sources shifted in the dispersion direction. Finally, it is possible to perform an optimal extraction with a complex extended background emission.

The \texttt{AdOpt} plugin is released with the \texttt{SMART} package (versions equal or later than 8.0). The code is available at the Infrared Science Center website, along with extensive documentation. \texttt{SMART} and the optimal extraction will be maintained in the future, with a possible inclusion of the optimal extraction for the high-resolution modules.

\acknowledgments We thank especially N.\ Chitrakar, D.\ G.\ Whelan, D.\ Levitan, and M.\ Devost for previous iterations of the optimal extraction code on which the present one is partly based. We also thank C.\ Tayrien and the disk team at Rochester for discussions on the spectral traces. Finally, we are grateful to D.\ Devost and J.\ R.\ Houck who initiated the idea of developing an optimal extraction for the IRS, and to the ISC team at Cornell for testing and support. We appreciate the help and comments of Fred Lahuis on this paper.

\end{document}